\documentclass[12pt]{article}
\usepackage{epsfig}

\begin{document}

\title{Staggering behavior of the low lying excited states of even-even nuclei in a
Sp(4,R) classification scheme}
\author{S.Drenska$^{\dag }$, A. Georgieva$^{\dag }$ and N.Minkov $^{\dag \ddag }$\\
\\
$^{\dag }$ Institute of Nuclear Research and Nuclear Energy,\\
Bulgarian Academy of Sciences, Sofia 1784, Bulgaria\\
$^{\ddag }$ RCNP Osaka University, 10-1 Mihogaoka, \\
Ibaraki city, Osaka 567-0047, Japan\\
}
\maketitle

\begin{abstract}
We implement a high order discrete derivative analysis of the low
lying collective energies of even-even nuclei with respect to the
total number of valence nucleon pairs $N$ in the framework of
$F$- spin multiplets appearing in an algebraic $sp(4,R)$
classification scheme. We find that for the nuclei
of any given $F$- multiplet the respective experimental energies exhibit a $%
\Delta N=2$ staggering behavior and for the nuclei of two united neighboring
$F$- multiplets well pronounced $\Delta N=1$ staggering patterns are
observed. Those effects have been reproduced successfully through a
generalized $sp(4,R)$ model energy expression and explained in terms of the
step-like changes in collective modes within the $F$- multiplets and the
alternation of the $F$-spin projection in the united neighboring multiplets.
On this basis we suggest that the observed $\Delta N=2$ and $\Delta N=1$
staggering effects carry detailed information about the respective
systematic manifestation of both high order $\alpha $ particle- like
quartetting of nucleons and proton (neutron) pairing interaction in nuclei.

PACS number(s):21.10.Re, 21.60.Fw
\end{abstract}

\section{Introduction}

The systematic behavior of the low-lying collective states of even-even
nuclei has been extensively studied in different phenomenological and
microscopic model approaches \cite{rcc, Bcl, bucl, bala}. In particular,
some algebraic classification schemes provide a convenient way to
systematize the general collective characteristics of these states in terms
of the basic nuclear structure quantities, such as the nuclear mass number,
the proton or neutron numbers, as well as the total number of neutron and
proton pairs in nuclear valence shells \cite{rcc,dggr,pvio}. These
quantities and their combinations can naturally be treated as the quantum
numbers of the irreducible representations associated with the respective
groups of symmetry. In this framework the low lying collective energy levels
exhibit various kinds of smooth systematic dependence on the different
quantum numbers, which can be evaluated empirically \cite{rcc,dggr}.

From another point of view, however, the current studies of various fine
effects in the structure of collective interactions and the respective
energy spectra suppose the presence of quite a complicated behavior of
nuclear collective characteristics. Thus, the application of the discrete
approximation of a high order derivative of given nuclear characteristics as
a function of particular physical quantity reveals various kinds of
staggering effects which carry detailed information about the fine
properties of nuclear interactions and the respective high order
correlations in the collective dynamics of the system.

Typical examples are the $\Delta I=1$ staggering effects (with $I$ being the
angular momentum) in the collective $\gamma $- bands \cite{BM75,MDRRB99} and
nuclear octupole bands \cite{BDDMRR1}, the $\Delta I=1,\,2$ and $4$,
staggering in superdeformed nuclear bands \cite{Fli,Ced,Stag,WZ97} and the $%
\Delta I=2$ staggering in the ground-state bands (gsb) of normally deformed
nuclei \cite{WT97}.

Another example is the odd-even staggering in nuclear binding energies \cite
{HiT,DOBNAZ,nbes} or the behavior of the nuclear masses \cite{sdn98} as a
function of the atomic number $A$, for which the influence of different
pairing, $\alpha-$like quartetting and mean-field effects are considered to
play a role.

In the above aspect, a natural question is whether the application of a high
order discrete derivative analysis to the low lying collective states in
wide range of nuclear chart would provide any staggering behavior of the
respective energies as functions of the quantum numbers of an appropriate
classification scheme. The answer of this question could open a new fine
tool in the interpretation and the systematic study of nuclear collective
interactions and their symmetries.

The purpose of the present work is to examine the above problem with respect
to the low lying collective states of even-even nuclei in the framework of
an algebraic $Sp(4,R)$ classification scheme \cite{girr}. The basic
assumption of this scheme is that the structure of nuclear proton and
neutron valence shells can be characterized in a unique way by the quantum
numbers of the irreducible representations (irreps) of the algebra $su(2)$
and the ladder representations of the algebra $u(1,1)$ which are subalgebras
of \ the boson representation of $Sp(4,R),$ considered as a classification
group. As a result the even-even nuclei of each major (proton and neutron)
shell can be classified into sets of $F$-spin multiplets \cite{pvio}, where
any nucleus is uniquely identified by the total number of valence bosons $N$
(pairs of protons and neutrons) and the third projection $F_{0}$ of the $F$%
-spin. It is known that in each $F$-spin multiplet the energies of the low
lying yrast states exhibit a smooth and periodic behavior as functions of $N$%
. On this basis, a unified theoretical description of gsb energies in
even-even nuclei has been obtained \cite{dggr}.

In the present paper, we apply the high order discrete derivative analysis
to the both experimental and theoretical $2_{1}^{+}$ energies as functions
of the quantum number $N$ in the framework of the respective $F$- multiplets
appearing in the $Sp(4,R)$ scheme. The same analysis is also applied to the
next higher $4^{+}$ and $6^{+}$ \ levels of the yrast bands of the
classified nuclei and some typical results are presented. As it will be seen
the experimental data exhibit considerable $\Delta N=2$ staggering behavior
in all considered $F$-multiplets (with $N$ even or odd) and also very well
pronounced $\Delta N=1$ staggering effect in the cases when two neighboring $%
F$- multiplets are united into one single multiplet (with both even and odd $%
N$). In addition, we extend the study in some isotopic chains, where a
tedious $\Delta N=1$ staggering effect can be observed also but its general
magnitude is quite small compared to the patterns observed in the $F$-spin
multiplets.

We will see that these phenomena are reproduced theoretically by a
generalized phenomenological energy expression derived in the framework of
the classification scheme. On this basis we are able to provide an adequate
physical interpretations of the observed staggering patterns in terms of the
related collective model characteristics and the high order $\alpha $-like
quartetting and pairing nucleon interactions as well.

In Sec. 2 we briefly outline the algebraic base of the $Sp(4,R)$
classification scheme. The high order discrete derivative
formalism and its meaning in the context of the present theory is
given in Sec. 3. Experimental and theoretical staggering patterns
with $\Delta N=2$ in the $F$ - multiplets and with $\Delta N=1$
in the combined (neighboring) $F$ - multiplets and isotopic
chains are presented in Sec. 4. A detailed analysis of their form
and amplitudes in the different multiplets is also given.\ In
Sec.5 the obtained results are analyzed in terms of the model
characteristics of the classification scheme. Some comments and
conclusions are given in Sec. 6.

\section{Low-lying collective states in the $Sp(4,R)$ classification scheme}

The nuclear $Sp(4,R)$ classification scheme is based on the use of two kinds
of bosons, proton $(\pi )$ \ and neutron $(\nu )$ bosons, interpreted as
pairs of the respective nucleons, confined in the valence orbits. The boson
representation of the algebra $sp(4,R)$ \ is generated through the
respective creation and annihilation operators as follows \cite{dggr}:

\begin{equation}
\begin{tabular}{llllll}
$\pi ^{\dagger }\pi ^{\dagger },$ & $\nu ^{\dagger }\nu ^{\dagger },$ & $\pi
^{\dagger }\nu ^{\dagger },$ & $\pi \nu ,$ & $\nu \nu ,$ & $\pi \pi ,$ \\
$N_{\pi }=\pi ^{\dagger }\pi ,$ & $N_{\nu }=\nu ^{\dagger }\nu ,$ &  &  & $%
F_{+}=\pi ^{\dagger }\nu ,$ & $F_{-}=\nu ^{\dagger }\pi .$%
\end{tabular}
\label{spge}
\end{equation}

The following subset of the above operators (\ref{spge}) : \
\begin{equation}
\begin{tabular}{ll}
$F_{+},$ & $F_{-},$%
\end{tabular}
F_{0}=\frac{1}{2}(N_{\pi }-N_{\nu })  \label{suf2g}
\end{equation}
generates a $su_{F}(2)$ algebra so that a $F-$spin quantum number is
assigned to the considered boson system with a mathematical structure
similar to that of the isospin.

The operator $N=N_{\pi }+N_{\nu }$ \ \ commutes with all the operators (\ref
{suf2g}) acting in this way as a first order invariant of an $u(2)\supset $ $%
su_{F}(2)$ $\otimes u_{N}(1)$ \cite{pvio,brsp}. The space $\mathrm{H,}$ in
which the boson representation of $sp(4,R)$ acts is reducible and the
operator $(-1)^{N}$ splits it into two irreducible spaces $\mathrm{H}_{+}$
and $\mathrm{H}_{-},$ corresponding to $N$ even or odd respectively. Acting
on these spaces the operator of the total number of particles $N$ reduces
them into a direct sum of a totally symmetric unitary irreps of $su_{F}(2),$
labeled by $N=0,2,4,...$ for $\mathrm{H}_{+}$ and $N=$ $1,2,3...$ for $%
\mathrm{H}_{-}$ $.$ The operator $F_{0}$ (\ref{suf2g}) does not differ
essentially from the first order Casimir operator of the noncompact $%
u(1,1)\subset sp(4,R).$ Hence it reduces the spaces $\mathrm{H}_{+}$ and $%
\mathrm{H}_{-}$ to the ladder series defined by its fixed eigenvalues. The
states from a given ladder, are defined by the eigenvalue of $F_{0}$ and are
obtained by the subsequent action of the operator $\pi ^{\dagger }\nu
^{\dagger }$ on the lowest weight state. The operator $F_{0}$ reduces each $%
u(2)-$ representation (with fixed value of $N$) to the representations of $%
u_{\pi }(1)\oplus u_{\nu }(1)$ labeled by $N_{\pi }$ and $N_{\nu }$ (\ref
{spge}) respectively. The same is obtained by reducing the $u(1,1)$ ladders
with the operator $N$.

In terms of the well known and rather successful language of IBM-2 \cite
{ibm2}, the algebraic operators (\ref{spge}) obtain the physical meaning of
basic nuclear characteristics inherent for the $Sp(4,R)$ classification
scheme. Thus the eigenvalues of the operators $N_{\pi }=\frac{1}{2}
(N_{p}-N_{p}^{(1)})$ and $N_{\nu }$ $=\frac{1}{2}(N_{n}-N_{n}^{(1)})$ are
naturally interpreted as the numbers of valence proton and neutron pairs for
a nucleus of a given shell. Here $N_{p}^{(1)}$ and $N_{n}^{(1)}$ are the
proton and neutron numbers respectively of the double magic nucleus at the
beginning of the shell, considered as the ``vacuum state'' for the
symplectic representations. A basic point in the present classification
scheme is that the major shells, which are defined by their bordering magic
numbers $(N_{p}^{(1)},N_{n}^{(1)}|N_{p}^{(2)}$,$N_{n}^{(2)})$ (with $%
N_{p}^{(2)}>N_{p}^{(1)}$ and $N_{n}^{(2)}>N_{n}^{(1)}$) are mapped on two $%
Sp(4,R)$- multiplets $(N_{p}^{(1)},N_{n}^{(1)}|N_{p}^{(2)}$,$%
N_{n}^{(2)})_{\pm }$, with $N$ being even $(+)$ or odd $(-)$. In these terms
the total number of valence bosons, $N$, and the third projection of the $F$%
-spin, $F_{0}$, are exactly the operators that reduce the $sp(4,R)$ \ spaces
to a definite vector in one of the $u_{\pi }(1)\oplus u_{\nu }(1)$
subspaces, corresponding to a given nucleus with fixed $N$ and $F_{0}$. In
this way the nuclei of given major shell are naturally classified into $%
F_{0} $ - multiplets (with fixed $F_{0}$ and $N$ taking all its possible
values \cite{pvio}), unified in each of the symplectic spaces $\mathrm{H}%
_{+} $ and $\mathrm{H}_{-}.$ The nuclei belonging to each of these
multiplets differ by $\Delta N=2$, an $\alpha $- like cluster (two protons
and two neutrons) created by the operator $\pi ^{\dagger }\nu ^{\dagger }$(%
\ref{spge}). It is important to remark that another kind \ of $\Delta N=2$
nuclear sequences \cite{dgi} can be defined in the symplectic spaces $H_{+}$%
\ and $H_{-}$ by the action of the operators $\pi ^{\dagger }\pi ^{\dagger }$%
\ (or $\nu ^{\dagger }\nu ^{\dagger }$)\ on the minimal weight
states of each $F_{0} $\ - multiplet. These sequences correspond
to the even-even nuclei from a  given isotonic (or isotopic
chain),which differ by four protons (or neutrons) with $F_{0}$
differing by $\Delta F_{0}=+1$ (or $-1$).

Furthermore, if the anticommutation relations of the boson
creation and annihilation operators $\pi ^{\dagger },\pi $ \ and
$\nu ^{\dagger },\nu $ \cite{susp} are formally introduced, one
can easily see that $sp(4,R)$ is the even part of a superalgebra,
which has the single boson operators as odd generators. These
operators practically relate the even and odd symplectic
multiplets. On this basis any two neighboring $F_{0}$- multiplets
from $\mathrm{H}_{+}$ and $\mathrm{H}_{-}$, can be united into
one single odd--even $F_{0}$- multiplet in which the neighboring
nuclei are determined
by the subsequent alternative action of the operators $\pi ^{\dagger }$($%
\Delta F_{0}=+\frac{1}{2})$ and $\nu ^{\dagger }$ ($\Delta F_{0}=-\frac{1}{2}%
)$\ and thus differ alternatively by $\Delta F_{0}=\pm \frac{1}{2}$. In
addition the nuclei belonging to such odd-even $F_{0}$- multiplets differ by
$\Delta N=1$. In the same way the isotonic and isotopic nuclear chains of
even even nuclei with $\Delta N=1$ can be obtained by \ the respective
consecutive action of $\pi ^{\dagger }$($\Delta F_{0}=+\frac{1}{2})$ \ and $%
\nu ^{\dagger }$ ($\Delta F_{0}=-\frac{1}{2})$ without a change
in the sign of $\Delta F_{0}$.

The above formalism has been applied for classifying the even-even nuclei
into symplectic spaces mapped on the respective major nuclear shells \cite
{girr}. Moreover, it provides a natural systematic of the yrast energy bands
observed in these nuclei. It has been shown that the low-lying yrast
energies exhibit a smooth and periodic behavior in each $F_{0}$ - multiplet,
which allows their unified description in terms of the nuclear collective
characteristics inherent for the $Sp(4,R)$ classification scheme. On this
basis the following generalized energy expression has been proposed \cite
{dggr}:
\begin{equation}
E_{yrast}(h_{k},I,\omega )=\alpha \{h_{k}\}I(I+\omega ).  \label{enom}
\end{equation}
The second term of Eq.~(\ref{enom}) represents the generalized collective
interaction characterized by the geometrical parameter $\omega .$ The latter
has the physical meaning of a measure for the interplay between the
vibrational ($\omega >20)$ \ and the rotational ($\omega =1)$ collective
modes (\ref{enom}) and reflects the respective changes in nuclear shape.
Since in heavy nuclei the shape is changed from almost spherical at the
beginning of given shell to axially deformed in the midshell region and back
towards spherical at its closure, the parameter $\omega $ changes
respectively from $\omega >20$ to $\omega =1$ and then again to $\omega >1$.
The values of $\omega $ have been determined \cite{dggr} for each of the
classified nuclei by using the experimental yrast energy ratios $\frac{%
E_{I+2}}{E_{I}},$ which are known to provide a reliable criteria \cite{cer}
of nuclear collectivity. So, the assigned integer values of $\omega $%
\ for each of the classified nuclei (fixed values of $N$\ and
$F_{0}$) \ determine its type of collectivity.

The first term in Eq.~(\ref{enom}), $\alpha \{h_{k}\}$, is a dynamical
(inertial) coefficient determined as a function of the six quantum numbers $%
h_{k},$ $k=1,2...,6$, which are specific for each nucleus in a
given shell. Generally, the variables $h_{k}$ depend on the
number of protons $N_{p}$ and neutrons $N_{n}$ and the four magic
numbers $N_{p}^{1}$, $N_{n}^{1}$, $%
N_{p}^{2}$, $N,_{n}^{2}$ and can be expressed in terms of the classification
quantum numbers $N$ and $F_{0}$. Thus for a fixed nuclear shell the inertial
parameter is given by the expression:
\begin{equation}
\alpha
\{h_{k}\}=A_{1}+A_{2}N+A_{3}F_{0}+A_{4}N^{2}+A_{5}F_{0}^{2}+A_{6}NF_{0},
\label{nip}
\end{equation}
where $A_{i},i=1,2,...6$ are phenomenological parameters determined overall
for all nuclear shells \cite{dggr}.

In general, the geometrical parameter $\omega $ has its specific
values for each of the classified nuclei . The inertial parameter
$\alpha \{h_{k}\}$ is a function only of $N$ in each $F_{0}$ -
multiplets, and of the two classification numbers in the combined
odd-even and isotopic multiplets.

\section{High order discrete derivative analysis of collective energies in
the $Sp(4,R)$ scheme}

As it became clear from Sec. 2, the even-even nuclei of given $F_{0}$-
multiplet appearing in the considered $Sp(4,R)$ classification scheme are
uniquely determined by the subsequent even (or odd) values of the valence
boson number $N$. This allows us to apply the high order discrete derivative
analysis to the low \ lying gsb energies $E(N)$ for a given angular momentum
$I=2,4,6$ in any considered $F_{0}$- multiplet, as a function of the quantum
number $N$, by analogy to the staggering analysis of rotational band
energies applied in terms of the angular momentum $I$ \cite
{MDRRB99,BDDMRR1,BDDMRR2}.

Since for any given $F_{0}$- multiplet the values of the quantum number $N$
of the neighboring members differ by $\Delta N=2$ ($\alpha -$particle), the
following quantity can be introduced:
\begin{eqnarray}
Stg(2N) &=&\frac{1}{16}(6\Delta _{2}E(N)-4\Delta _{2}E(N-2)-4\Delta
_{2}E(N+2)  \nonumber \\
&&+\Delta _{2}E(N+4)+\Delta _{2}E(N-4)),  \label{stag}
\end{eqnarray}
where
\begin{equation}  \label{eq:stagen}
\Delta _{2}E(N)=E(N+2)-E(N).
\end{equation}
The point function $Stg(2N)$ is proportional to the discrete approximation
of the fourth derivative of the function $\Delta _{2}E(N)$, and obviously,
to the fifth derivative of the energy $E(N)$:
\begin{eqnarray}
Stg(2N) &=&\frac{1}{32}(10E(N+2)+5E(N-2)+E(N+6)  \nonumber \\
&-&\left[ 10E(N)+5E(N+4)+E(N-4)\right] ).
\end{eqnarray}
As such, it naturally obtains zero values for any polynomial form of $E(N)$
of power less or equal to four. Therefore, any non-zero values of the
quantity $Stg(2N)$ will imply a higher order functional dependence.
Moreover, any staggering (zigzagging) behavior of the $Stg(2N)$- values will
suggest the presence of quite complicated non-polynomial dependence of the
lowest collective energy and the respective nuclear interactions on the
quantum number $N$.

On the above basis it is expected that the application of the quantity $%
Stg(2N)$ to the low-lying collective nuclear states within the $F_{0}$%
-multiplets of $Sp(4,R)$ could provide a detailed information about the
possible influence of $\alpha $-particle-like high order quartetting
interaction ($\pi ^{\dagger }\nu ^{\dagger }\pi \nu $) of nucleons on the
systematic behavior of collective excitation energies.

We should remark, that the application of the function $Stg(2N)$
for the energies of the nuclei in the isotopic and isotonic
chains is restricted, since even in the larger shells there is
not enough number of points (observed excitation energies).
Actually this is due to the circumstance that the available data
are separated in the two spaces $\mathrm{H}_{+}$ and
$\mathrm{H}_{-}.$The above \ restriction is naturally released in
the extended regions of exotic nuclei (such as with $N_{n}=N_{p}$
and close to the drip lines), where the number of newly obtained
experimental data grows continuously. Now, we can extend the
framework of the fine systematic analysis of low-lying collective
energies by considering the unified odd-even $F_{0}$ -
multiplets. Within these multiplets the neighboring members
differ by $\Delta N=1$ (a proton or neutron pair) in a way that
the increasing $N$ corresponds to a \emph{subsequent alternative}
adding of proton and neutron pairs to the valence shells. In this
case the systematic
behavior of the excitation energies can be characterized analogously to Eq. (%
\ref{stag}) by a discrete approximation of the fourth derivative of the
energy difference for $\Delta N=1$:
\begin{eqnarray}
Stg(1N) &=&\frac{1}{16}(6\Delta _{1}E(N)-4\Delta _{1}E(N-1)-4\Delta
_{1}E(N+1)  \nonumber \\
&&+\Delta E_{1}(N+2)+\Delta _{1}E(N-2)),  \label{stag1}
\end{eqnarray}
where
\begin{equation}
\Delta _{1}E(N)=E(N+1)-E(N).  \label{edn1}
\end{equation}

The use of the same order in the discrete derivative approximations, (%
\ref{stag}) and (\ref{stag1}), for the $F_{0}$-multiplets with
$\Delta N=2$ and the combined odd-even and isotopic multiplets
with $\Delta N=1$ respectively, allows one to treat the $\alpha
$- particle quartetting and the different kinds of nucleon
pairing correlations in nuclei on the same footing, as well as to
compare quantitatively their fine systematic influence on nuclear
collectivity. Also, in this way the present classification scheme
allows the standard analysis of the $\Delta N=1$ staggering
effect in isotopic chains of even even nuclei. In the isotonic
chains this analysis is limited due to the Coulomb restriction on
the possible proton numbers, but nevertheless it seems to be
perspective in view of the current progress in the experimental
investigations of proton rich nuclei.

\section{$\Delta N=2$ and $\Delta N=1$ staggering patterns for the low lying
collective energies. Experiment and theory}

We have applied the functions $Stg(2N)$ (\ref{stag})  and
$Stg(1N)$ (\ref{stag1}) for the analysis of the experimental data
\cite {expe1,expe2,expe3} on the lowest collective energies of
even-even nuclei classified in the respective symplectic
multiplets. In all the considered cases these functions are
oscillating in respect to the zero of their scale, with a
changing amplitude, i.e. exhibit staggering patterns, which will
be analyzed bellow.

The function (\ref{stag}) has been used in all possible cases, in which the $F_{0}$%
-multiplets of the considered $Sp(4,R)$ classification scheme \cite{dggr},
contain more than 8 nuclei. For all of them we observe $\Delta N=2$
staggering behavior of the $2_{1}^{+}$ energy. Some typical staggering
patterns including either odd or even $F_{0}$-multiplets are shown on
\textbf{Fig. 1.}

\begin{figure}[th]
\caption{The experimental values of $Stg(2N)$, (\ref{stag}), for
the $F_{0}$ - multiplets: (a) $F_{0}=3/2$ from
$(28,50|50,82)_{-}$; (b) $F_{0}=-4$ from
$(50,50|82,82)_{+}$; (c) $F_{0}=0,1,2$ from $(50,82|82,126)_{+}$; (d) $%
F_{0}=1$ from $(50,82|82,126)_{+}$ for $I=2,4,6$ .}
\label{fig:1}
{}
\par
\epsfysize=10cm \centerline{\hbox{%
\epsfig{figure=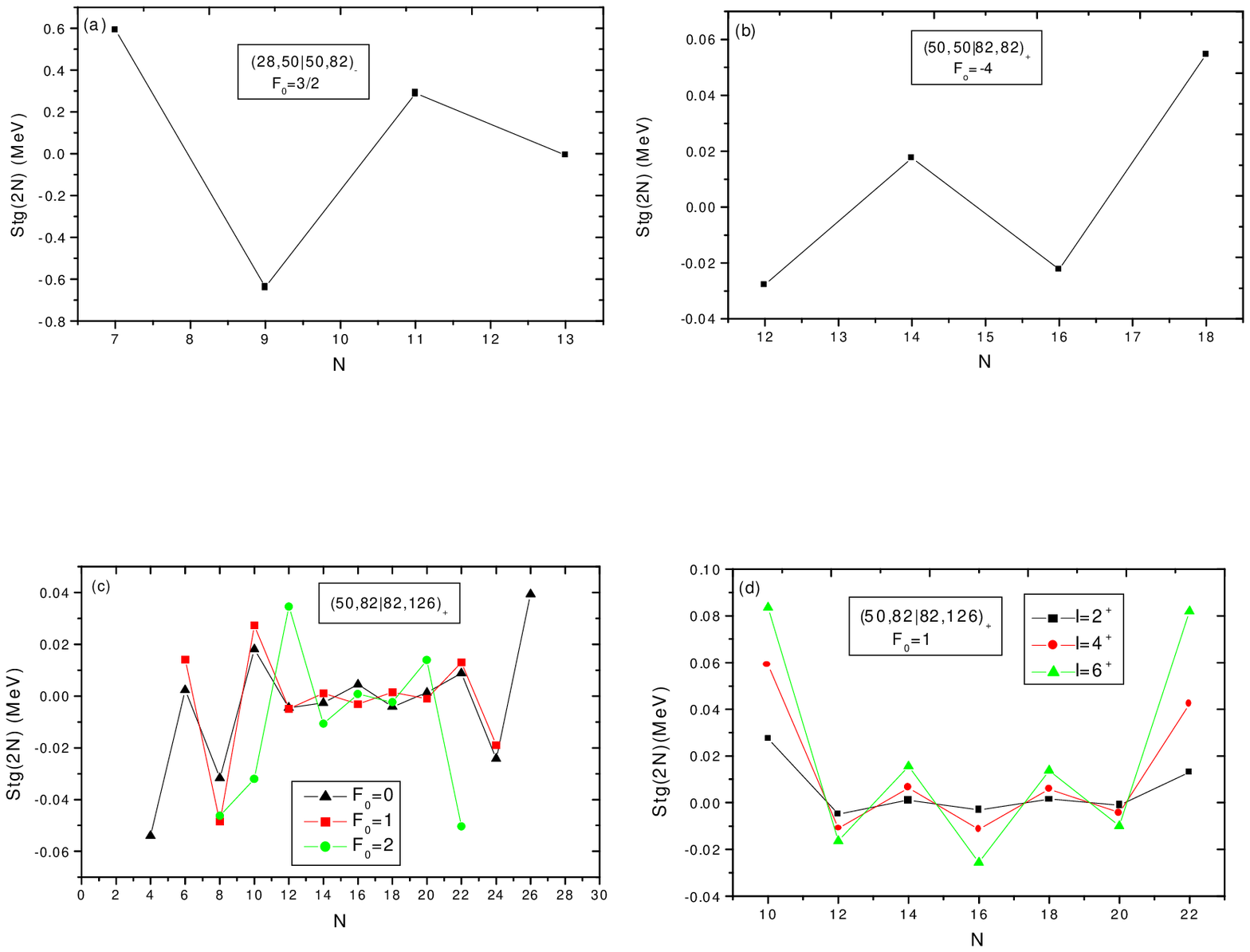,width=9cm,height=8cm}}} 
\end{figure}

We see, that for the different $F_{0}$-multiplets the general
scale of the staggering amplitude depends essentially on the
particular valence
shells. Thus, for the two multiplets $F_{0}=3/2$ \textbf{(Fig. 1(a))} and $%
F_{0}=0,1,2$ \textbf{(Fig. 1(c))}, belonging to the essentially different
(light and heavy) valence shell configurations $(28,50|50,82)_{-}$ and $%
(50,82|82,126)_{+}$ respectively, the staggering amplitude generally differs
by more than one order in magnitude. While for the lighter shells $%
(28,50|50,82)_{-}$ \ the function $Stg(2N)$ oscillates between $-0.8$ and $%
0.6$ MeV \textbf{(Fig. 1(a))}, for the heavier (rare earth) shells
it drops in the interval $[-0.06,0.04]$ MeV \textbf{(Fig. 1(c))}.
Some intermediate staggering magnitude in the interval
$[-0.03,0.05]$ MeV is observed for the multiplet $F_{0}=-4$
\textbf{(Fig. 1(b)),} which belongs to the ''intermediate''
valence space $(50,50|82,82)_{+}$.

As a result, we find that the general $\Delta N=2$ staggering scale tends to
be large for the light nuclear valence shells and essentially smaller, by
more than one order in magnitude, for the heavy nuclear regions. However, we
should remark that the more detailed comparison is restricted due to the
fact that the ''length'' of the $F_{0}$- multiplets increases towards the
heavier valence shells and, as it will be pointed out below, more
complicated structure of the staggering patterns appears there.

So, in the region of heavy nuclei we find that for any particular $F_{0}$
-multiplet the $\Delta N=2$ staggering function strongly varies in
dependence on the quantum number $N$. On \textbf{Fig. 1(c)} we show three
well developed (long enough) patterns obtained for the multiplets $%
F_{0}=0,1,2$ of the rare earth valence shells $(50,82|82,126)_{+}$. We find
that in the beginning of any multiplet $(N=4-12)$ as well as near its end $%
(N>20)$ a considerable staggering amplitudes of about $0.05$ MeV are
observed, while in the middle of the multiplet $(N=15-20)$, which
corresponds to the effective middle of the valence shells, the staggering
effect is essentially suppressed to practically zero amplitude. Also, we
remark that in some cases (see the multiplet $F_{0}=2$ on \textbf{Fig. 1(c)}%
)the transitions between the beginning, middle and the end of the multiplet
are characterized by irregularities in the zigzagging curve with subsequent
changes in the phase of the staggering pattern. This is a specific
indication for some kind of phase transition in nuclear collectivity within
the particular multiplet.

Further, in \textbf{Fig. 1(d)} we illustrate the $\Delta N=2$ staggering
patterns obtained for the experimental levels with angular momentum $I=2,4,6$
within the long $F_{0}=1$\ multiplet. The different curves are characterized
with the same phases of the oscillation, while the amplitude increases with
the increasing of the angular momentum, conserving the same dependence on
the quantum number $N$.

Now, let us consider the fine structure of the odd--even $F_{0}-$ multiplets
obtained by uniting two neighboring $F_{0}$-multiplets from the spaces $%
H_{+} $\ and $H_{-}$. The respective $2_{1}^{+}$- energy sequences are
obtained by switching between the subsequent members of the two considered
multiplets. The unified odd-even $F_{0}-$ multiplets contain much larger
number of nuclei than the single multiplets, which allows one to extend the
analysis by applying the discrete function ~(\ref{stag1}) to all the nuclear
shells, including the first $(28,28|50,50)_{\pm }$\ and the last (open) $%
(82,126|126,...)_{\pm }$\ major shells. In all of them we found that the
experimental energies exhibit rather well pronounced $\Delta N=1$\
staggering patterns. Some typical examples illustrating different
combinations of odd-even multiplets from all the shells are shown in \textbf{%
Fig. 2.}

First of all, we see that the general scale of the $\Delta N=1$\
staggering is larger by almost one order of magnitude than the
scale in the $\Delta N=2$ \ staggering patterns of the respective
separate $F_{0}$-multiplets. (For example, compare \textbf{Fig. 2
(c)} and \textbf{Fig. 1(c) }.)

Further, we find that the trend of the decrease of \ the staggering
amplitude with the increase of the shell dimension is observed with no
exceptions in this case, like in the $\Delta N=2$\ case. This can be seen,
for the odd-even multiplets with $F_{0}=\{-5/2,-2\}\ $from the shell $%
(28,28|50,50)_{\pm }$\ on \textbf{Fig. 2(a)} and from the shell $%
(82,126|126,...)_{\pm }$\ \ \cite{expe3} on \textbf{Fig. 2(d)}, where the
maximal staggering amplitude drops from $0.2MeV$\ in the first one, to $%
0.003MeV$\ in the last one. This decrease is gradual through the consecutive
shells \textbf{( Fig. 2(a) -(d)).}

\begin{figure}[tbp]
\caption{ The experimental values of $Stg(1N)$, (\ref{stag1}), for
the odd-even $F_{0}$ - multiplets: (a) the odd-even multiplet
$F_{0}=\{-5/2,-2\}$ from $(28,28|50,50)$; (b) the odd-even
multiplet $F_{0}=\{1,3/2\}$ from  $(28,50|50,82)$; (c) the
odd-even multiplet $F_{0}=\{1,3/2\}$ from $(50,82|82,126)$; (d)
the odd-even multiplet $F_{0}=\{-5/2,-2\}$ from
$(82,126|126,...)$; (e) the odd-even multiplet
$F_{0}=\{-4,-7/2\}$ from $(50,50|82,82)$ for $I=2,4,6$ .}
\label{fig:2} {}
\par
\centerline{\hbox{\epsfig{figure=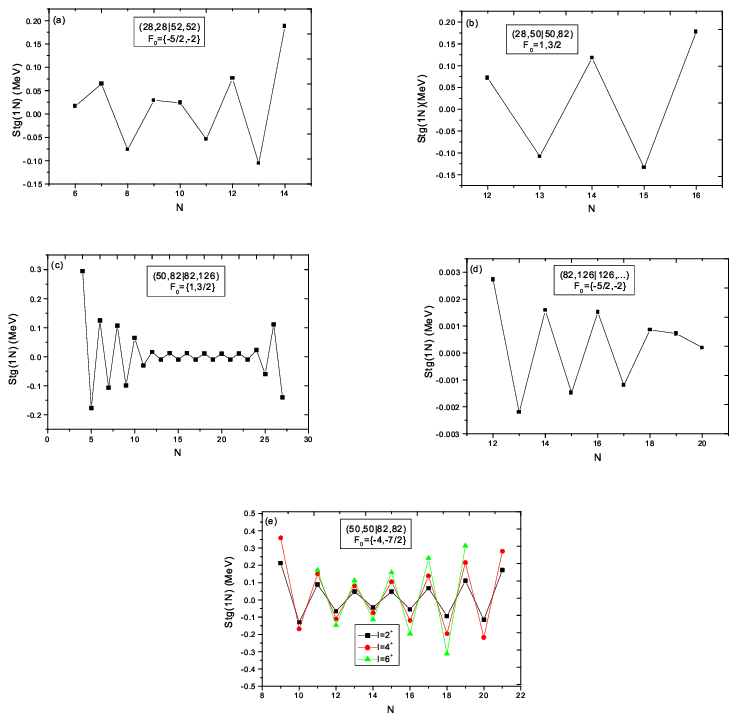,width=14cm,height=15cm}}}
\end{figure}

As in the $\Delta N=2$\ case, we observe that in the beginning and the end
of any particular unified odd--even multiplets the staggering amplitude is
larger, than in the middle of it. It is however important to remark that in
the middle of the unified $F_{0}-$multiplets in the ''lighter''shells $%
(28,28|50,50)_{\pm }$, $(28,50|50,82)_{\pm }$ \ and $\
(50,50|82,82)_{\pm }$ the decrease in the staggering magnitude is
not very well expressed, \ as could be seen for the multiplets
$F_{0}=\{-5/2;-2\}$ on \textbf{Fig. 2(a)}, $F_{0}=\{1,3/2\}$ on
\textbf{Fig. 2(b)} and  $F_{0}=\{-4;-7/2\}$ on \textbf{Fig.
2(e)}$\}$\ $\ $respectively. \ In the middle of all the combined
$F_{0}-$multiplets in the rare - earth $(50,82|82,126)_{\pm }$\
region the staggering amplitude is almost vanishing, as illustrated for the $%
F_{0}=\{1,3/2\}$ \ multiplet on \textbf{Fig. 2(c).}\

The application of ~(\ref{stag1})\textbf{\ }to the energies of the next
exited states $I=4,6$\ in the combined odd-even multiplets, shows as in the $%
\Delta N=2$ staggering, that all the typical features of the energy behavior
of the $2_{1}^{+}$-states, outlined above are even enhanced with the
increase of the angular momentum. An example for one of the longest combined
multiplets $F_{0}=\{-4;-7/2\},$ from the shell $(50,50|82,82)_{\pm }$ is
given on \textbf{Fig. 2(e).}

The nuclear isotonic chains obtained in the scheme are too short , so\textbf{%
\ }the $Stg(1N)$\ \ function is investigated only in the isotopic chains of
even- even nuclei, obtained by the consecutive action of the boson operators
$\nu ^{+}$ (pairs of neutrons) . It does not exhibit a pronounced staggering
effect. An exception is presented on \textbf{Fig. 3}, for the $Os$ isotopes,
but the \ maximal amplitude in this case is two orders smaller $(0.006MeV)$
\ than the odd-even $\Delta N=1$\ staggering in the same shell\textbf{\
(Fig. 2(c)).}

\begin{figure}[bh]
\caption{ The experimental values of $Stg(1N)$, (\ref{stag1}) as function of
the neutron numbers $N_{n}$, obtained for the $Os$ isotopic multiplet from $%
(50,82|82,126)$. }
\label{fig:3}\centerline{\hbox{%
\epsfig{figure=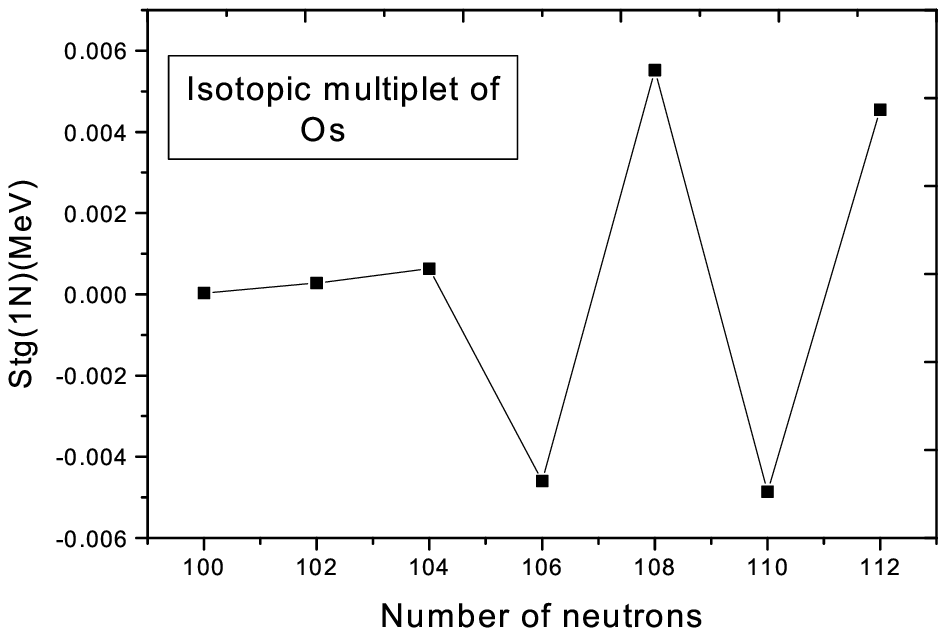,width=5cm,height=4cm}}} 
\end{figure}

Now, we turn to the analysis of above phenomena in terms of the
collective model characteristics inherent for the applied
$Sp(4,R)$ classification 
of the theoretical energies (\ref{enom}) are obtained as
functions of coefficients $\alpha \{h_{k}\}$ and through them on
the classification numbers $N$, $F_{0}$ (\ref{nip}) and the
relevant value of $\omega $ for each nucleus.
\begin{figure}[ph]
\caption{ The experimental and theoretical values (\ref{enom}) of
the functions: (a) $Stg(2N)$, (\ref{eq:stagen}) for the multiplet
$F_{0}=0$ from $(50,82|82,126)_{+}$ ; (b) $Stg(1N)$,
(\ref{stag1}) for the odd-even multiplet $F_{0}=\{0,1/2\}$  from
$(50,82|82,126)$ ; (c) same as (b) for the odd-even multiplet
$F_{0}=\{1,1/2\}$ from $(50,82|82,126)$ for $I=2$; (d) same as
(c) for $I=4$; (e) same as (c) for $I=6$. }
\label{fig:4}\epsfysize=12cm \centerline{\hbox{%
\epsfig{figure=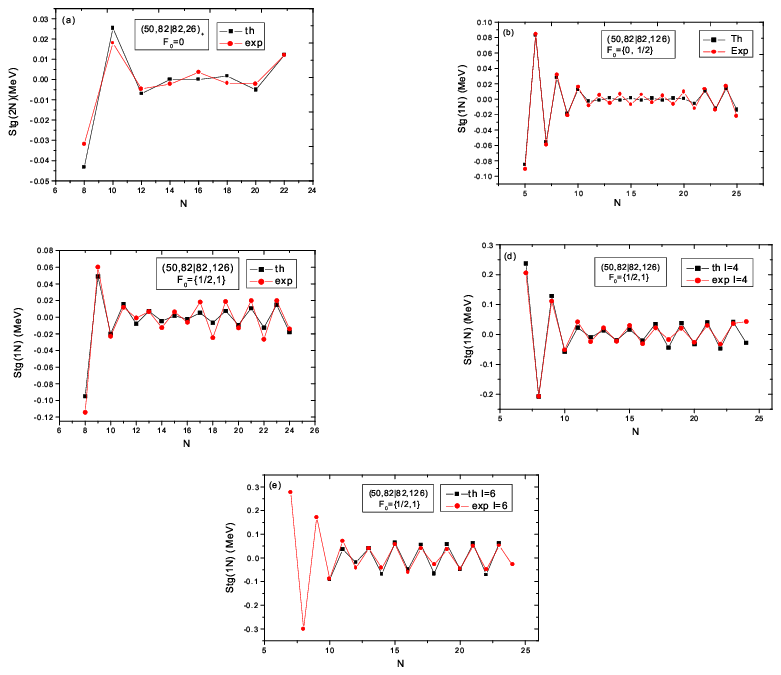,width=15cm,height=15cm}}} 
\end{figure}
Thus, the theoretical energies $E_{yrast}(h_{k},I,\omega )$
successfully reproduce the respective experimental data
\cite{dggr} allowing us to apply our fine systematic analysis by
situating them into the considered multiplets of the
classification.

Hence, we apply the staggering filters (\ref{stag}) and
(\ref{stag1}) on the theoretically obtained by means of
(\ref{enom}) energies of the low lying collective states of
even--even nuclei in all the multiplets, in which experimental
$\Delta N=2$ and $\Delta N=1$ staggering was observed. We found
that the behavior of the (\ref{stag}) and (\ref{stag1}) is
reproduced rather well by the theoretical energies. This result
is illustrated for $2_{1}^{+}$ energies from the multiplet
$F_{0}=0$ of the shell $(50,82|82,126)_{+}$ and for the odd-even
$F_{0}=\{0,1/2\}$ \ multiplet of the same shell \ on \textbf{Fig.
4(a)} and \textbf{Fig. 4(b)}, respectively. We see that in both
cases with $\Delta N=2$ and $\Delta N=1$ \ the specific systematic
dependence of the staggering magnitude on the quantum number $N$
is reproduced. Moreover in the ends of the multiplets the larger
amplitudes are reproduced together with the correct signs. The
theoretical amplitude is a bit smaller than the experimental one
and its irregularity is more clear in the region of the well
deformed nuclei. As illustrated on \textbf{Fig. 4(c), 4(d)
}and\textbf{\ 4(e) }the staggering behaviour \ of the energies of
the states with $I=2,4,6$ respectively in the combined multiplet $%
F_{0}=\{1/2,1\} $ from the shell $(50,82|82,126)$ is reproduced
by the generalized formula (\ref{enom}), including the sign and
typical features of the amplitude for the higher states. This
means, that all the presented in this work analysis for the first
\ exited $2_{1}^{+}$ states is appropriate for all the higher
yrast states, which were included in the fitting procedure, when
(\ref{enom}) was deduced in \cite{dggr}. The slight discrepancies
in the theoretical reproduction of the staggering amplitudes are
due to the accuracy of the phenomenological description and the
restriction to integer values of $\omega .$

\section{Model analysis}

\bigskip A relevant model interpretation of the presented investigation, can
be obtained on the basis of the geometrical factor $\omega ,$ non
explicitly depending on $N$  and $F_{0},$ defining the nuclei in
the multiplet. Its phenomenological values \cite{dggr} vary in the limits $%
\omega =1\div 30$ and specify the character of the collective
motion. In decreasing order these values consequently characterize
the collective excitations of nearly spherical, vibrational
nuclei (large values of $\omega \approx 15-30;U(5)-$limit of
IBM1); the transitions region (with fast decrease of $\omega $)
including the $\gamma -$soft nuclei with $\omega \approx 5$;
(O(6) - limit) and the region of well deformed, rotational nuclei
($\omega =1;SU(3)$). In the lighter shells \textbf{(Fig. 1(a),
1(b), 2(a), 2(e)) } we have a transition only from spherical to
$\gamma -$soft nuclei and back to almost the same value of
$\omega ,$ which corresponds to the transition from the U(5) to
the O(6) limit of the IBM1 \cite{ibmb} (only one side of the
Casten's triangle \cite{catri}). \ The typical model distribution
of $2_{1}^{+}$ energy levels between the different $\omega $-
modes of collective motion is illustrated on \textbf{Fig. 5} for the long $%
F_{0}=0$ multiplet of the shell $(50,82|82,126)_{+}$. There the
thin curves represent various $\omega $-fixed modes of the \
$E_{yrast}(h_{k},2,\omega )$ as functions of the valence pairs
number $N,$ while the thick curve connects its phenomenologically
determined values at a specific $\omega$. We see that in the
beginning of the multiplet ($N=4$) the theoretical $2_{1}^{+}$
energies start from the high $\omega =17$ ''vibrational'' curve ,
but with the increase of $N$ they rapidly dump down (through
$\omega =8,6$) to the ''rotational iso-line''
with $\omega =1$ which ''receives'' all levels in the middle region $N=10-22$%
. Further, in the end of the multiplet (shell closure region $N=24-30$)
several jumps up to the transitional ($\omega =3,6,8$) collective modes are
observed.

Namely, the above\textbf{\ }rather large discrete jumps of energy
as a function of $N$ \ between the different $\omega $- lines
cause the pronounced $\Delta N=2$ staggering effect in the
beginning and the end of the theoretically ''filled''
$F_{0}$-multiplet. The small (but non zero) staggering amplitudes
appearing in the middle region can be considered as the effect of
the five-point discrete derivative, (\ref{eq:stagen}), which has a
''memory'' of the preceding states and also ''propagates'' the
information about ''what happens'' in the ends.

\begin{figure}[tp]
\caption{ The theoretical values $E_{yrast}(N,\protect\omega )$
(\ref{enom}) for different values of the parameter
$\protect\omega $ and the real behavior of the energies (thick
line) for the specific $\protect\omega $
for the multiplet $F_{0}=0$ from the shell $(50,82|82,126).$ }%
\epsfysize=5cm \centerline {\epsfbox{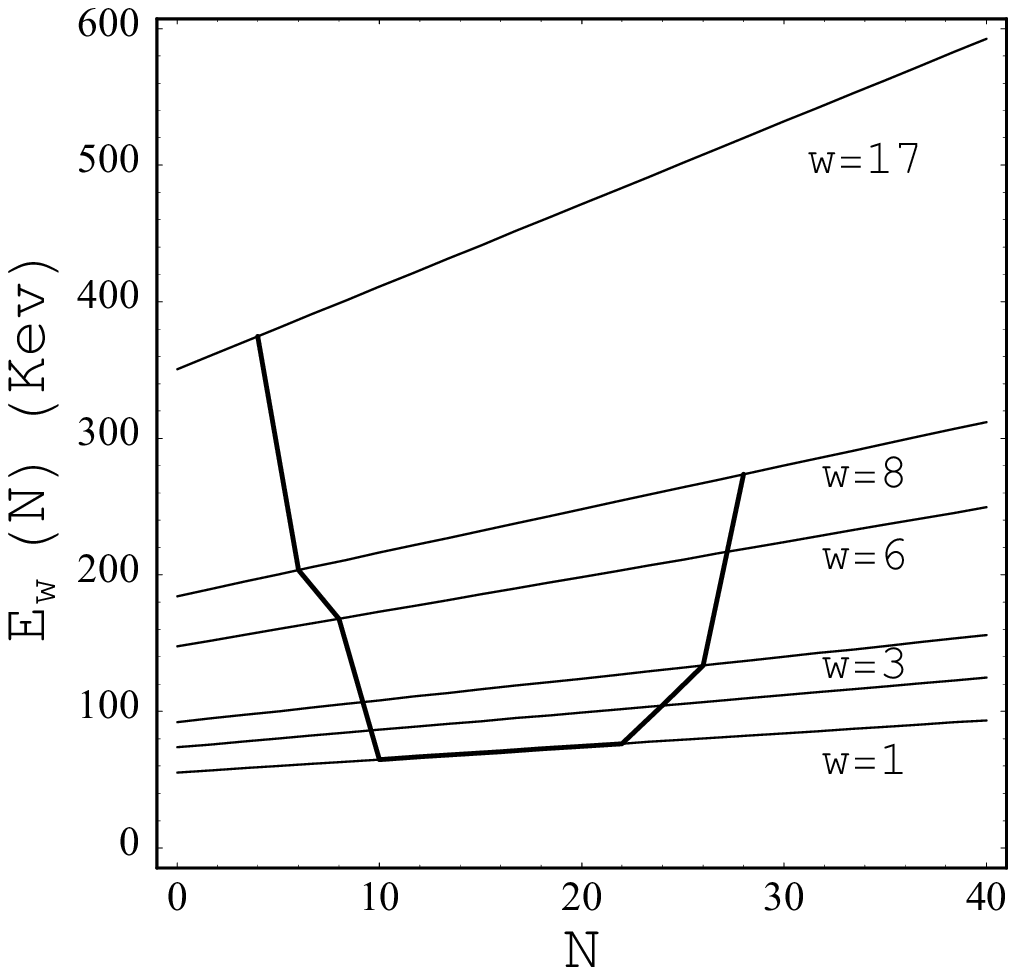}} \label{fig:5}
\end{figure}

At the same time, the change in the geometric factor $\omega ,$
trough its dependence on $N$, reflects phenomenologically the
respective change in the intrinsic nuclear structure and its
influence on the character of collective motion. On this basis we
can directly interpret the experimental $\Delta N=2$ staggering
patterns as the result of rapid change in nuclear collective
properties with respect to different intrinsically
determined modes, related in this case to the accumulation of $%
\alpha -$like clusters.

Now, we can directly extend the above considerations with respect to the $%
\Delta N=1$ staggering patterns corresponding to changes in the
intrinsic nuclear structure with alternating kinds of nucleon
pairs. In the combined symplectic multiplets the analysis
concerning the geometric factor $\omega $ and its contribution to
the fine systematical behavior of $2_{1}^{+}$ energies is still
valid for their members, but their more gradual change must be
taken into account. It provides an analogous explanation of the
large $\Delta N=1$ staggering amplitudes in the ends of given
odd-even multiplet. However, in this case there is another factor
that appears to be decisive especially for the essentially larger
staggering scale compared to the $\Delta N=2$ case. In the single
$F_{0}$ multiplet the difference between two neighboring energies
is entirely determined by the quantum number $N$ and the
geometric factor $\omega $ (all other model quantities and
parameters are fixed). In the combined multiplets there is
additional dependence on the quantum number $F_{0}$ (\ref{nip}).
The
alternative change of the $F_{0}$- values with $+\frac{1}{2}$ then $-\frac{1%
}{2}$ in the two neighboring $F_{0}$- multiplets provides respective change
in the inertial factors \ $\alpha \{h_{k}\}$ (\ref{nip}) and thus a'priori
produces a mutual energy oscillation in the respective energy sequences.
\textbf{\ }Hence, even in the energy difference (\ref{edn1}) some staggering
is obtained, due to its dependence on \ the oscillating values of $F_{0}$ as
seen on\textbf{\ \textbf{Fig. 6}.}

\begin{figure}[tbp]
\caption{ The experimental values of $\Delta _{1}E(N)$, (\ref{edn1}) for the
odd-even $F_{0}=\{1/2,1\}$ - multiplet from the shell $(50,82|82,126).$ }
\label{fig:6}\epsfysize=12cm \centerline{\hbox{%
\epsfig{figure=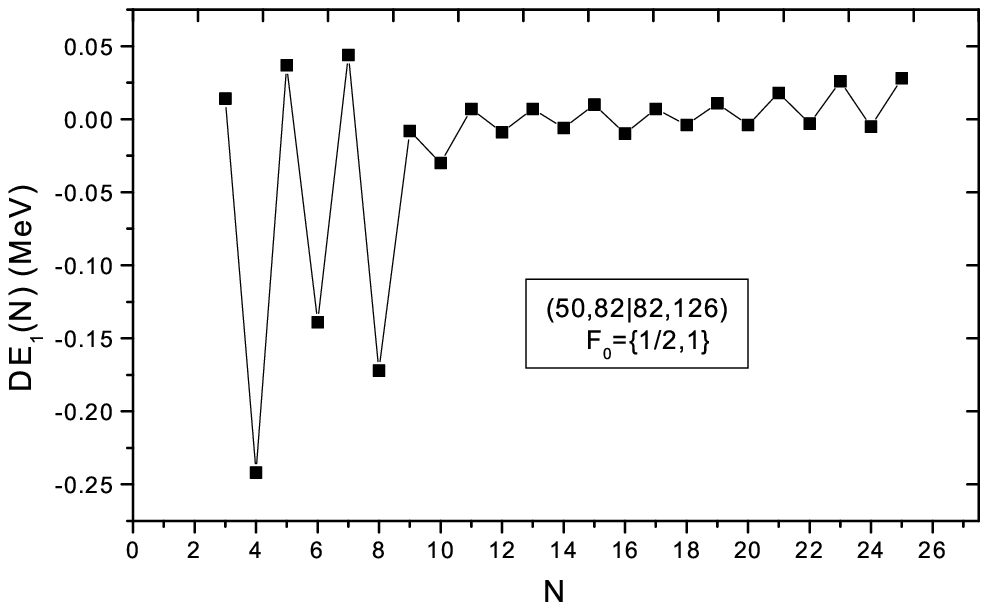,width=5cm,height=4cm}}} 
\end{figure}

The latter results in magnifying the $\Delta N=1$ staggering background.
Thus, our theoretical analysis suggest that the form and the sensitivity of $%
\Delta N=1$ staggering effect to the nuclear shell structure are determined
by discrete changes of the quantum number $\omega $ in respect to $N$, but
its general magnitude is also determined by the fluctuations of the $F_{0}-$%
values. The magnitude and the form of the staggering pattern, together with
the respective set of phenomenological $\omega $-values provide both
relevant qualitative and quantitative characteristics of nuclear
collectivity in the framework or the $Sp(4,R)$ classification scheme.

\section{Conclusions}

The results presented so far allow us to outline the following systematic
behavior of the experimentally observed low lying collective energy levels
in the framework of the considered $Sp(4,R)$ model scheme. The energies of
the levels classified in any single $F_{0}$-multiplet of $Sp(4,R)$ exhibit $%
\Delta N=2$ staggering behavior, while the levels of more general odd-even
-multiplets provide respective $\Delta N=1$ staggering patterns of their
fifth order discrete derivative. In both cases the general staggering
magnitude is relatively large for the light nuclear valence shells and
considerably smaller for the valence shells of heavy nuclei. In addition,
both $\Delta N=2$ and $\Delta N=1$ staggering patterns of heavy nuclei are
characterized by relatively large amplitudes in the ends and their
suppression in the middle region of the respective single and combined
odd-even $F_{0}$-multiplets . On the other hand, we find that the $\Delta
N=1 $ staggering effect is by an order of magnitude \emph{stronger} and more
\emph{regular} in form than the $\Delta N=2$ staggering. For the odd-even $%
F_{0}$-multiplets the staggering is inherent even for the energy differences
of the neighboring states (\ref{edn1}), with $F_{0}$ integer and half
integer. The weakest staggering effect is observed in the isotopic nuclear
chains, where there is no oscillation of the $\Delta F_{0}$ values$.$

The application of the $Sp(4,R)$ classification scheme provides relevant
physical interpretation of the experimentally observed $\Delta N=2$ and $%
\Delta N=1$ staggering effects in terms of the nucleon correlations in
nuclear valence shells. The $\Delta N=2$ effect indicates very fine
systematic behavior of nuclear collectivity associated with the \emph{%
simultaneous} adding of proton and neutron pairs ($\alpha $-like \ clusters)
to valence shells, relevant to the effect of four nucleon interaction. In
the $\Delta N=1$ staggering effect the situation is different. Here the
filling of valence shells is realized in two ways :

\begin{enumerate}
\item  in the odd-even multiplets by the alternative adding of proton and
neutron pairs, \ which is reflected by the alternative change of the $F_{0}-$%
values and thus strongly magnifies the staggering effect,

\item  in the isotopic multiplets by adding of only neutron pairs, which
gives a very small staggering.
\end{enumerate}

In these two cases the observed effect is due to the pairing correlations,
which are of \ lower order, but its magnitude strongly depends on the charge
fluctuations in the consecutive adding of the nucleon pairs.

This analysis provides an indication of the different ways of influence of
the different orders of valence nucleon correlations (pairing and
quartetting) on the fine systematic characteristics of the nuclear
collectivity.

The general formula for the yrast energies \ $E_{yrast}(h_{k},I,\omega )$
obtained for the even-even nuclei, as classified in the symplectic
multiplets reproduces rather accurately the experimental $\Delta N=2$ and $%
\Delta N=1$ staggering. . The geometric parameter $\omega $, introduced in
\cite{dggr}, not only unifies the description of the energies of the ground
state bands of the spherical, transitional and well-deformed nuclei, but
also is the main reason for obtaining the effects of the shape transitions
reflected in both staggering effects. \ The amplitude of the staggering
generally follows the same trend as $\omega $, so it could be considered as
relevant characteristic of the type of collectivity\textbf{\ }of nuclear
spectra and respectively the nuclear shape. \textbf{\ }In the region of the
well deformed nuclei, when $\omega $ $=1$ the \ staggering effect is
strongly suppressed, which is observed in the rare -earth and actinide
shells and is the reason for the larger scale in the light ones.

The generalizations of the interaction coefficient \ $\alpha \{h_{k}\}$ (\ref
{nip}), as functions of the classification quantum numbers, case is
important for the reproduction of the staggering effects. In the $\Delta N=2$
case $F_{0}$ is fixed, as well as in the isotopic multiplets with $\Delta
N=1 $, while in the combined odd-even multiplets its values oscillate from a
state to state. This reflects in the bigger values of the amplitudes and
their regularity in the $Stg(1N)$\textbf{\ \ }function in the combined
multiplets.

The investigation in this work, reveals once again that the symplectic
classification scheme is rather convenient for a generalization and
reproduction of important physical properties of the even -even nuclei. It
also has a predicting power, since in the consideration of the staggering $%
\Delta N=1$ effects in the last shell, we used experimental data \cite{expe3}%
, which was not included in the fitting procedure in \cite{dggr}. As the
ladder series of the boson representations are infinite \ dimensional, the
symplectic multiplets can be extended with the newly obtained data for
exotic nuclei and they can be considered as a new test for this kind of fine
structure investigations. Moreover the suggested quantitative analysis could
provide a rather fine estimation of the possible energy regions, where low
lying collective excitations of exotic nuclei can be expected.

In conclusion, the above results show that the investigation of the
different types of the staggering effects based on a convenient
classification scheme is an useful tool to understand not only the
collective properties of the lowest states of even--even nuclei , but also
provides various quantitative considerations on the fine structure of the
changes in nuclear collectivity.

\textbf{Acknowledgments:}

The authors are grateful to Prof. D.Bonatsos for useful
discussions of this work. One of the authors (AG) is grateful to
Prof. J.P. Draayer for the hospitality in Louisiana State
University and the interest of all the group on the subject of
the paper. This work has been supported by the Bulgarian National
Fund for Scientific Research under contract no MU--F--02/98.

\end{document}